\documentclass[12pt]{article}
\usepackage{a4}
\usepackage{subfigure}
\usepackage{graphicx}

\begin{document}
\title{An accurate analytic solution to the Thomas-Fermi equation}
\author{M. Turkyilmazoglu}
\maketitle
\begin {center}
{\large\emph{Mathematics Department, University of Hacettepe,
06532-Beytepe, Ankara, Turkey}}
\end {center}

\renewcommand{\baselinestretch}{1.}
\begin{abstract}
The explicit analytic solution of the Thomas–-Fermi equation
thorough a new kind of analytic technique, namely the homotopy
analysis method, was employed by Liao \cite{Liao2003a} (Appl.
Math. Comp. 144, (2003)). However, the base functions and the
auxiliary linear differential operator chosen were such that the
convergence to the exact solution was fairly slow. New base
functions and auxiliary linear operator to form a better homotopy
are the main concern of the present paper. It is known that proper
choice of base functions and auxiliary operator is extremely
significant in gaining the exact solution in order to reduce the
computational cost. The proposed homotopy here not only greatly
reduces the computational efforts by at least doubling the
convergence of the homotopy series, but also enlarges the
convergence region of the homotopy series as compared with that of
Liao \cite{Liao2003a}. Pad\'e approximants to the obtained
solutions increase the accuracy even to a higher degree. To
support this, the explicit analytical expressions obtained using
the proposed approach are compared with the numerically computed
ones and those of Liao \cite{Liao2003a}.
\end{abstract}

\section{Introduction}
\label{introduction} Nonlinear phenomena are encountered in all
areas of sciences and engineering. Even though the study of
nonlinear equations is of great importance to many scientific
researchers in various fields, it is very difficult to solve
nonlinear problems and, in general, it is often more costly to get
an analytic approximation than a numerical one to a given
nonlinear problem \cite{Liao2003a}.

Since the Thomas-Fermi equation is one typical not easy-to-find
exact solution, the explorations of the solution have taken
considerable attraction of many authors. The analytic
approximations of the Thomas-Fermi equation were proposed by some
different techniques such as the variational approach
\cite{Burrows84}, the $\delta$-expansion method
\cite{Laurenzi90,Cedillo93}, the decomposition method
\cite{Chan87,Wazwaz99}, see also \cite{Civan84,Allan92,Pert99}.
Since the techniques used in these papers were mostly perturbative
and limited, the solutions are obliged to be valid for certain
regions with a decreasing accuracy.

In an aim to search for new and more effective analytical tools,
Liao in 1992 \cite{Liao92,Liao92a} proposed the homotopy analysis
method which deforms a difficult nonlinear problem into easy
linear counterparts. This is achieved by introducing an auxiliary
parameter in the construction of an homotopy, which can provide a
convenient way to control the convergence of the approximation
series and adjust the convergence regions if necessary. A series
of nonlinear problems were later attacked with the use of homotopy
analysis method
\cite{Liao95,Liao99,Liao2001,Liao2003,Liao2004a,Liao2006} and
\cite{3Turkyilmazoglu2009,4Turkyilmazoglu2009}. An elegant, simple
and explicit analytic solution to the Thomas-Fermi equation was
presented in \cite{Liao2003a}. One main drawback of the homotopy
analysis method is that it's convergence to the exact solution
becomes too time-consuming as the parameters involved in the
equations are not treated rationally. This seems to be the reason
why Liao in \cite{Liao2003a} was unable to obtain higher-order
approximations for the solution of the Thomas-Fermi equation, but
presented only the homotopy approximations with a percentage
accuracy of at most first order. On the other hand, this order is
inadequate to demonstrate convergence of the method to the true
solution.

We in the present paper use homotopy analysis technique of Liao
for the analytic calculation of Thomas-Fermi equation. Motivated
by the study \cite{Liao2003a}, we aim to show that the
computational task during the implementation of the homotopy
analysis method can be further reduced if the Thomas-Fermi
equation is treated in a more rational manner. This is
accomplished here by modifying the original approach of Liao
\cite{Liao2003a}. Knowing that the choice of base functions and
auxiliary linear differential operator is very important in the
homotopy technique, see for example \cite{Liao2003book}, new base
functions and auxiliary linear differential operator are presented
here. Unlike those of \cite{Liao2003a}, using these new parameters
it is shown here that the convergence of the homotopy can be
greatly accelerated, even greater by the homotopy Pad\'e
approximants. In addition to this, the region of convergence is
much improved with the present approach.

The following strategy is adopted in the rest of the paper. In \S
\ref{Method} the basis of homotopy analysis method is laid out
with an application to the nonlinear Thomas-Fermi equation.
Analytic expressions for the solution are derived and compared
with the numerical and previously published results in \S
\ref{Results}. Finally conclusions follow in \S \ref{conclusions}.

\section{The Homotopy Analysis Method}
\label{Method} The Homotopy analysis method was first proposed by
the Chinese mathematician Liao \cite{Liao92a}. This method is
based on the homotopy and has several advantages. To underline,
firstly its validity does not depend upon whether or not nonlinear
equations under consideration contain small or large parameters,
hence it can solve more of strongly nonlinear equations than the
perturbation techniques. Secondly, it provides us with a great
freedom to select proper auxiliary linear operators and initial
guesses so that uniformly valid approximations can be obtained.
Thirdly, it gives a family of approximations which are convergent
in a larger region. Liao successfully applied the homotopy
analysis method to solve some nonlinear problems in mechanics. For
example, Liao in \cite{Liao99} gave a purely analytic solution of
2D Blasius's viscous flow over a semi-infinite flat plate, which
is uniformly valid in the whole physical region.

The essential idea of this method is to introduce a homotopy
parameter, say $p$, which varies from 0 to 1. At $p=0$, the system
of equations usually has been reduced to a simplified form which
normally admits a rather simple solution. As $p$ gradually
increases continuously toward 1, the system goes through a
sequence of deformations, and the solution at each stage is close
to that at the previous stage of the deformation. Eventually at
$p=1$, the system takes the original form of the equation and the
final stage of the deformation gives the desired solution. Here we
apply it to solve the nonlinear Thomas-Fermi problem. Consider a
differential equation used to calculate the electrostatic
potential in the Thomas-Fermi atom model \cite{Fermi27,Thomas27},
called the Thomas-Fermi equation with the boundary conditions
\begin{eqnarray}\label{eq1}
   {{\rm d}^2u\over {\rm d}x^2}+\sqrt{u^3\over x} = 0,\;
   u(x=0)=1,\;u(x\to \infty)=0,
\end{eqnarray}which describes the spherically symmetric charge
distribution about a many electron atom. Equation (\ref{eq1}) is
not amenable to exact treatment and, therefore, approximate
techniques must be resorted to. There exists neither linear terms
nor small or large parameters in equation (\ref{eq1}), so the
standard perturbation methods cannot be applied directly. Due to
the fact that the homotopy analysis method requires neither a
small parameter nor a linear term in a differential equation, one
possibility to approximately solve equation (\ref{eq1}) is by
means of the homotopy analysis method.

We first rewrite the original equation (\ref{eq1}) in the form
\begin{eqnarray} \label{eq2}
   N(u) = x u''^2-u^3=0;\quad u(0)=1,\;u(\infty)=0.
\end{eqnarray}

The essence to approximate a problem is to represent its solution
by means of a complete set of base functions. Considering the
boundary conditions in (\ref{eq2}) and the physical meaning of
$u$, Liao \cite{Liao2003a} chosen the set of base functions
\begin{eqnarray}\label{eq3}
  \{(1+x)^{-m}|m\ge 1\}.
\end{eqnarray}

As Liao \cite{Liao2003book} discusses in detail on the free
falling problem, the selection of different sets of functions is
possible all of which generate the same solution. Besides, the
homotopy analysis method provides us with freedom to choose the
initial guess and the auxiliary linear operator so that one can
represent the solution of the Thomas-Fermi equation by distinct
set of base functions. However, some set of base functions might
cause a very slow convergence to the true solution. Therefore, if
a proper choice is made while selecting a set of function a better
convergence can be achieved which may save both time and resources
while computing. Thus, we here take a different set of base
functions from (\ref{eq3}) in an aim to speed up the convergence
to the exact solution and instead of (\ref{eq3}) choose
\begin{eqnarray}\label{eq3a}
\{\alpha (\alpha+\beta x)^{-\gamma m}|m\ge 1, \quad \alpha, \beta,
\gamma \in R\}
\end{eqnarray}to represent the solution $u(x)$ of (\ref{eq2}) in the
form
\begin{eqnarray}\label{eq4}
  u(x) &=& \sum_{n=1}^\infty c_n {\alpha \over (\alpha+\beta x)^{\gamma
  n}},
\end{eqnarray}where $c_m$'s are coefficients. Notice that if we
set $\alpha=\beta=\gamma=1$ in (\ref{eq3a}-\ref{eq4}), the base
functions of Liao (\cite{Liao2003a}) in equation (\ref{eq3}) are
obtained. Therefore, our base functions are somewhat more general.
The key to choose the parameters in (\ref{eq3a}) is that a better
convergence rate can be captures as compared to the case of Liao
\cite{Liao2003a}. It will be clear soon that some certain
selections will indeed lead to high savings in the computational
time of the approximate solution of Thomas-Fermi equation by
accelerating its convergence.

Equation (\ref{eq3a}) provides us with the {\em rule of solution
expression}. This rule is important in the frame of the homotopy
analysis method. Considering the initial conditions in (\ref{eq2})
and the {\em rule of solution expression} above described, it is
obvious that
\begin{equation}\label{eq5}
  u_0(x)={\alpha \over (\alpha+\beta x)^{\gamma}}
\end{equation}is a good initial guess for $u(x)$ satisfying the
boundary conditions in (\ref{eq2}) exactly. We next choose here
$$L={\beta x+\alpha\over \alpha+\gamma}{\partial^2\over \partial
x^2}+\beta {\partial \over \partial x}$$ as our auxiliary linear
differential operator, which was found to be quite efficient for
the consideration of the present nonlinear problem, which has the
property that $L(C_1(\alpha+\beta x)^{-\gamma}+C_2)=0,$ with $C_1$
and $C_2$ arbitrary constants. We are now at the stage of
constructing the zeroth order deformation equation system
associated with the Thomas-Fermi problem given by (\ref{eq2}),
which is also called a family of differential equations (viewing
$p$ as a parameter)
\begin{eqnarray}\label{eq6}
  (1-p)L(u) &=& p h N(u),\quad u(0,p)-1=u(\infty,p)=0.
\end{eqnarray}
It should be noted that the nonlinear differential operator in
equation (\ref{eq4}) is given as in (\ref{eq2}). Moreover, the
parameter $h$ is an auxiliary non-zero parameter to adjust the
convergence rate of the perturbation series, which was found to be
$-1/2\leq h<0$ in \cite{Liao2003a} for the convergence. However,
in our case we will show that the region of $h$ for the
convergence can be extended. Obviously when $p=0$ and $p=1$, we
have respectively
\begin{eqnarray}\label{eq7}
  u(x,0) = u_0(x), \quad u(x,1) = u(x).
\end{eqnarray}

Hence the process of giving an increment to $p$ from 0 to 1 is the
process of $u(x,p)$ varying continuously from the initial guess
$u_0(x)$ to the final solution $u(x)$. This kind of continuous
variation is called deformation in topology so that we call system
(\ref{eq6}) the zeroth order deformation equation. Next,
differentiating (\ref{eq6}) successively and eventually imposing
at $p=0$, the kth-order deformation equations follow as
\begin{eqnarray}\label{eq8}
  L(u_k) &=& L(u_{k-1})+hR_k,\quad u_k(0)=u_k(\infty)=0.
\end{eqnarray}
The function $R_k$ on the right-hand side of (\ref{eq8}) is given
by
\begin{eqnarray*}
    R_k &=& {1\over (k-1)!}{\partial^{k-1}N \over \partial
    p^{k-1}}|_{p=0}.
\end{eqnarray*}

Finally, a straightforward Taylor expansion of $u(x,p)$ at the
point $p=0$ and eventually imposing the series at $p=1$ gives the
solution of system (\ref{eq2}) in the form
\begin{eqnarray}\label{eq9}
  u(x) &=& u_0(x)+\sum_{k=1}^\infty u_k(x),
\end{eqnarray}
for which we presume that the initial guesses $u_0$ to $u$, the
auxiliary linear operator $L$ and the non-zero auxiliary parameter
$h$ are all so properly selected that the deformations $u(x,p)$
are smooth enough and their kth-order derivatives with respect to
$p$ in equation (\ref{eq8}) exist and are given by $u_k={1\over
k!}{\partial u\over \partial p}|_{p=0}$. It is clear that the
convergence of Taylor series at $p=1$ is a prior assumption here
so that the system in (\ref{eq9}) holds true. The formulae in
(\ref{eq9}) provide us with a direct relationship between the
initial guesses and the exact solutions. Moreover, a special
emphasize should be placed here that the kth-order deformation
system (\ref{eq8}) is a linear differential equation system with
the auxiliary linear operator $L$ whose fundamental solution is
known as aforementioned. Eventually, we obtain the result at the
Nth-order approximation as
\begin{eqnarray}\label{eq10}
  u(x) &=& \sum_{k=0}^N u_k(x).
\end{eqnarray}

\section{Results and discussion}\label{Results}
In this section analytic approximate solutions corresponding to
system (\ref{eq2}) is presented. Solutions obtained from the
homotopy analysis method are compared with those obtained from the
full numerical computations and the available solution of
\cite{Liao2003a}.

Note that the analytic solution presented in (\ref{eq10}) contains
the auxiliary parameter $h$, which can be employed to control the
convergence of approximations and adjust convergence regions when
necessary. As claimed before, the proper choice of parameter
$\alpha,\beta,\gamma$ can yield an extended convergence region as
compared to the case of Liao \cite{Liao2003a}.
\begin{figure}[!htb]
\begin{center}
\mbox{ \subfigure[]{
\includegraphics[width=7cm,height=7cm]{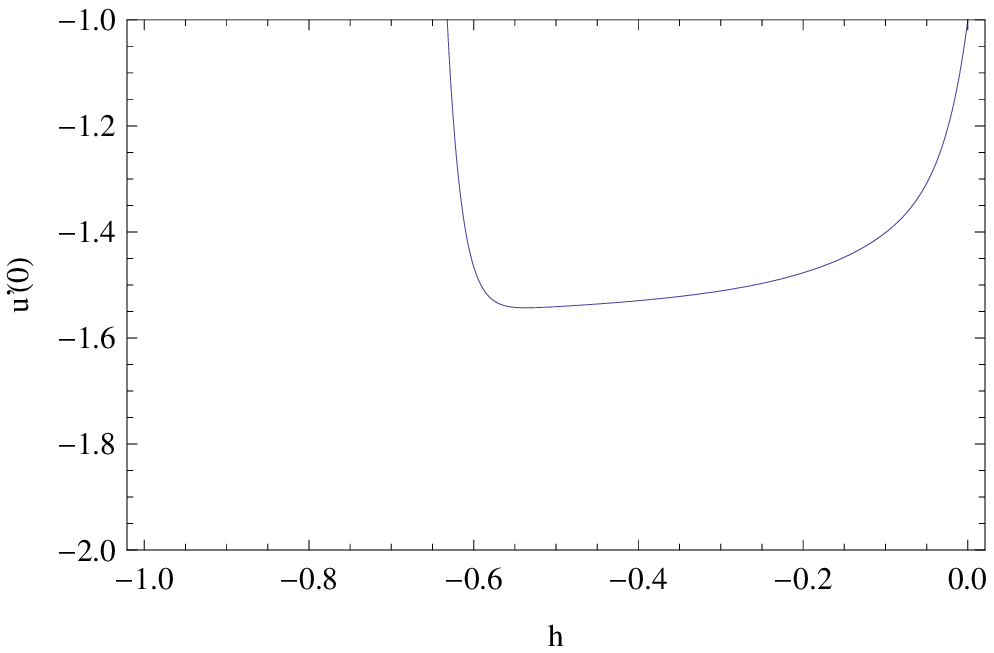}}
\subfigure[]{
\includegraphics[width=7cm,height=7cm]{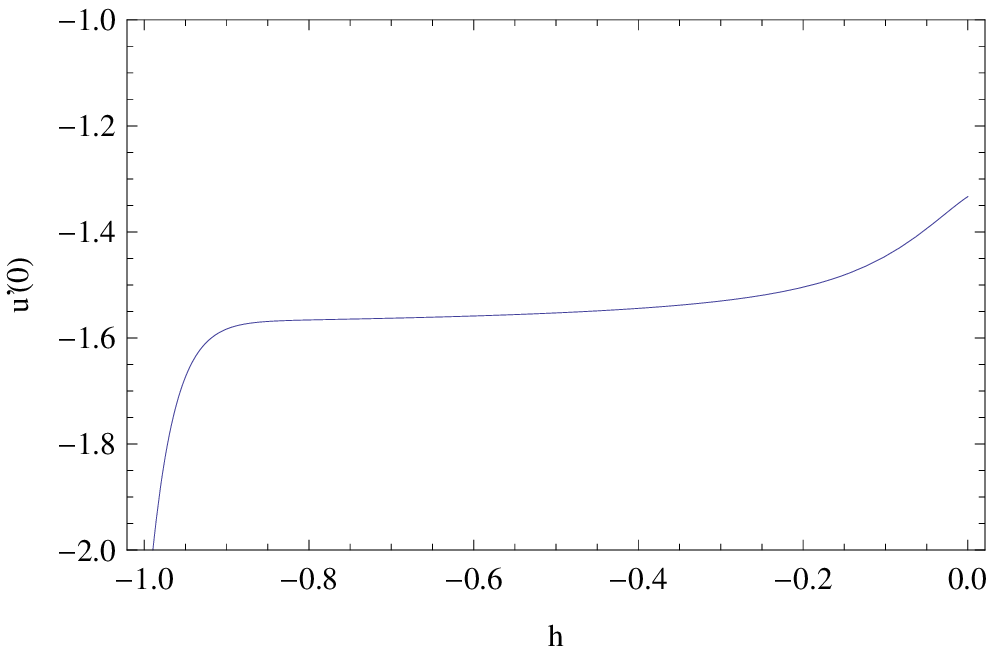}}}
\caption[Basic flow]{The $h-$ curves of $u'(0)$ obtained from the
20th-order homotopy approximation solution of the problem
(\ref{eq2}). (a) Case of Liao \cite{Liao2003a}
($\alpha=\beta=\gamma=1$), (b) the present case
($\alpha=3/4,\beta=\gamma=1$).} \label{fig0}
\end{center}
\end{figure}

To illustrate this, the influence of $h$ on the convergence of the
solution series are given in figures \ref{fig0}(a-b). The
$h-$curves are drawn for $u'(0)$ obtained from the 20th-order
homotopy analysis approximation. Figure \ref{fig0}(a) depicts the
case of \cite{Liao2003a} (not given in \cite{Liao2003a}) and
\ref{fig0}(b) depicts our case. As found by Liao in
\cite{Liao2003a}, when $h$ is restricted to $-1/2\leq h<0$, the
series in (\ref{eq10}) converges in the whole region $0\leq x<
\infty$, which is indeed the case as shown in figure
\ref{fig0}(a). On the other hand, for the parameters $\alpha=3/4$,
$\beta=\gamma=1$, we are able to extend the convergence region of
$h$ to $-9/10\leq h<0$, as shown in figure \ref{fig0}(b). This is
important since in a larger region of convergence a better
convergence rate can be achieved for the series. In fact, it is
easy to see that in order to have a good approximation $h$ has to
be chosen around $-4/5$.
\begin{figure}[!htb]
\begin{center}
\includegraphics[width=14cm,height=10cm]{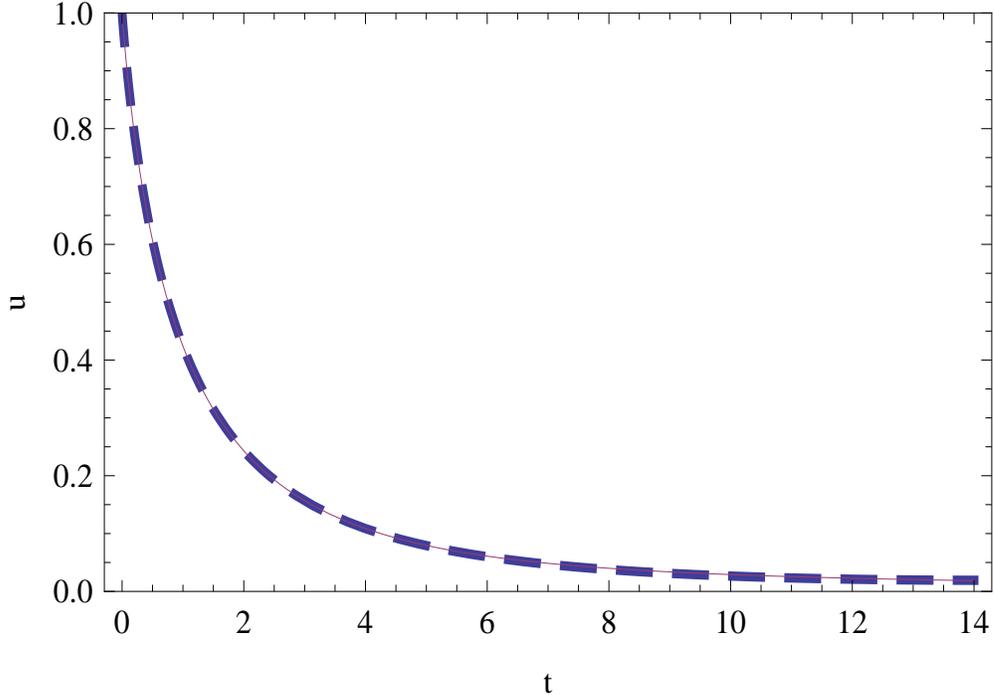}
\caption[Basic flow]{Comparison of the analytic result of the
Thomas-Fermi equation with the numerical result. Solid line
denotes analytic result at the 40th-order approximation for
$h=-4/5$ and dashes denote the numerical result.} \label{fig1}
\end{center}
\end{figure}
When $h=-4/5$, the analytic result at the 40th-order of
approximation agrees well with the numerical result, as shown in
figure \ref{fig1}. As compared with the 60th-order of
approximation of Liao \cite{Liao2003a} (see figure 2 of Liao
\cite{Liao2003a}) our analytic solution at a lower order is
clearly seen to better approach the exact solution. To explain the
reason deeply, and in fact to disclose the difference between our
solution and that of Liao \cite{Liao2003a}, we list in table
\ref{Table 1} the values of initial slope $u'(0)$ (which is used
to obtain the energy of a neutral atom in the Thomas–Fermi model,
see \cite{Fermi27}) and curvature $u''(0)$ respectively. The error
between the exact and approximate $u'(0)$ is also given in table
\ref{Table 1}. Obviously, the error decreases as the order of
approximation increases, in both our case and the case of Liao
\cite{Liao2003a}. However, more importantly, the convergence of
our series (\ref{eq10}) is seen to be more than twice as the
convergence of the iterations of Liao \cite{Liao2003a}. This
justifies the claim that our base functions indeed approximate the
exact solution of Thomas-Fermi equation at a better convergence
rate. This as a result will save us both in terms of computational
time and resources. Moreover, owing to equation (\ref{eq2}), it
holds $u''(0)\to \infty$ as $x\to 0^+$. The approximations of
$u''(0)$ obtained from the homotopy analysis method when $h=-3/4$
are listed in table \ref{Table 1} show that $u''(0)$ of the
analytic solution (\ref{eq2}) indeed tends to infinity. Again, the
trend is observed to be twice as high as the trend of Liao
\cite{Liao2003a}. The remarkable accuracy of the results tabulated
clearly illustrate the effectiveness and efficiency of the
proposed approach. Table further reveals as claimed before that
the presented approximate solutions take less computational time
as compared with that of \cite{Liao2003a}.
\begin{table}[!htb]
\begin{center}
\begin{tabular}{cccccc}
  \hline $N$  &$u'(0)$ & Error(\%) & $u'(0)^L$ & $u''(0)$ & $u''(0)^L$\\\hline
  $10$        & -1.54628 & 2.63 & -1.50014 & 25.4567 & 13.0003  \\
  $20$        & -1.56597 & 1.39 & -1.54093 & 46.8426 & 23.0819  \\
  $30$        & -1.57305 & 0.94 & -1.55595 & 68.1948 & 33.1119  \\
  $40$        & -1.57669 & 0.71 & -1.56373 & 89.5378 & 43.1275  \\
  $50$        & -1.57891 & 0.57 & -1.56848 & 110.877 & 53.1370  \\
  $60$        & -1.58040 & 0.48 & -1.57168 & 132.214 & 63.1434  \\
  $70$        & -1.58171 & 0.40 & -1.57399 & 151.216 & 73.1480  \\
  $80$        & -1.58303 & 0.31 & -1.57572 & 173.012 & 83.1514  \\
  $90$        & -1.58424 & 0.24 & -1.57708 & 196.871 & 93.1542  \\
  $100$       & -1.58515 & 0.18 & -1.57816 & 224.112 & 103.1560 \\
\end{tabular}
\end{center}
\caption{Analytic approximations of $u'(0)$ and $u''(0)$ when
$h=-3/4$, the percentage error occurred for $u'(0)$ and comparison
of the results with those of Liao \cite{Liao2003a}, denoted by a
superscript $L$.}\label{Table 1}
\end{table}

As implemented in \cite{Liao2003a}, we can further employ the
[m,m] diagonal homotopy Pad\'e approximants \cite{Baker75} to the
power series of $u'(0)$ in order to gain more accurate
approximations of the initial slope, as shown in table \ref{Table
2}. Note that the error decreases with the increase of the degree
of the Pad\'e approximants. Comparisons with the Pad\'e
approximations of Liao \cite{Liao2003a} once more shows the better
accuracy obtained from the present approach.
\begin{table}[!htb]
\begin{center}
\begin{tabular}{cccc}
  \hline Pad\'e approximants &$u'(0)$ & Error(\%) & $u'(0)^L$ \\\hline
  $[10,10]$  & -1.58030 & 0.48933 & -1.51508  \\
  $[20,20]$  & -1.58571 & 0.14867 & -1.58281  \\
  $[30,30]$  & -1.58694 & 0.07122 & -1.58606  \\
  $[40,40]$  & -1.58752 & 0.03469 & -1.58668  \\
  $[50,50]$  & -1.58801 & 0.00384 & -1.58712  \\
\end{tabular}
\end{center}
\caption{Approximations of the initial slope $u'(0)$ given by the
diagonal Pad\'e approximants, when $h=-3/4$ and the percentage
error occurred. Comparisons with $u'(0)$ obtained by Liao
\cite{Liao2003a} is denoted by the superscript $L$.}\label{Table
2}
\end{table}

\section{Concluding remarks}
\label{conclusions} In this paper, the homotopy analysis method
has been applied to obtain approximate analytical solution for
nonlinear phenomena governed by the Thomas-Fermi equation. The
results presented have readily revealed that the approach adopted
is very effective and convenient. Comparisons of the
approximations with the published ones of Liao \cite{Liao2003a}
have proven the accuracy and efficiency of the proposed approach
which rapidly doubles the convergence rate of the homotopy series
solution to the exact solution and hence significantly reduces the
time consumption while evaluating the approximate analytic
solutions thorough the homotopy analysis method.

The key to get fast convergence and better accuracy as compared
with that of Liao \cite{Liao2003a} has been to select more
appropriate different set of base functions and auxiliary linear
differential operator. Such a rational choice has been shown to
enlarge the region of convergence of the homotopy series and thus
yield acceleration of the convergence. Diagonal homotopy Pad\'e
approximants to the power series obtained have been shown to
improve some degree the accuracy and convergence of the homotopy
series to the exact ones.

The purely explicit analytical solutions obtained here also
provide a good scientific base for the validation of the
numerically computed values using different schemes in the
literature. In addition to this, the developed approach could be
expected to be applicable to attain the solutions of highly
nonlinear systems arising from the applications in engineering and
science.

\bibliographystyle{unsrt}
\bibliography{mybib}
\end{document}